\documentclass[]{JHEP3}

\usepackage{amssymb,amsmath,graphicx}
\usepackage[all]{xy}
\usepackage{dcolumn}
\usepackage{bm}
\usepackage{bbm}
\usepackage{amsmath,amssymb,amsthm}
\usepackage{mathrsfs}
\usepackage{slashed}
\usepackage{ucs}

\usepackage[utf8x]{inputenc}

\newcommand{\be}{\begin{equation}}
\newcommand{\ee}{\end{equation}}
\newcommand{\bea}{\begin{eqnarray}} 
\newcommand{\eea}{\end{eqnarray}}
\newcommand{\ft}[2]{{\textstyle\frac{#1}{#2}}}

\def\Re{\mathop{\rm Re}\nolimits}
\def\Im{\mathop{\rm Im}\nolimits}

\def\rme{{\mathrm e}}

\newsavebox{\uuunit}
\sbox{\uuunit}
    {\setlength{\unitlength}{0.825em}
     \begin{picture}(0.6,0.7)
        \thinlines
        \put(0,0){\line(1,0){0.5}}
        \put(0.15,0){\line(0,1){0.7}}
        \put(0.35,0){\line(0,1){0.8}}
       \multiput(0.3,0.8)(-0.04,-0.02){12}{\rule{0.5pt}{0.5pt}}
     \end {picture}}
\newcommand {\unity}{\mathord{\!\usebox{\uuunit}}}

\csname @addtoreset\endcsname{equation}{section}

 \def\cD{{\cal D}} 
  
  \def\cN{{\cal N}} 
    
 \def\cO{{\cal O}}

\title{$F$-term uplifting and the supersymmetric integration of heavy moduli}
\author{ A.~Achúcarro$^{1,2}$, S.~Hardeman$^{1}$ and K.~Sousa$^{1}$
\\
$^1$Lorentz Institute of Theoretical Physics, Leiden University, \\ 
Niels Bohrweg 2, 2333 RA Leiden, The Netherlands \\
$^2$ Department of
Theoretical Physics, University of the Basque Country UPV-EHU, \\ 
P.O. Box 644, 48080 Bilbao, Spain\\
Email:  \email{achucar@lorentz.leidenuniv.nl}, \email{sjoerd@lorentz.leidenuniv.nl}, \email{kepa@lorentz.leidenuniv.nl}}
 
\abstract {
  We study in detail the stability properties of the simplest $F$-term uplifting mechanism consistent with the integration of heavy moduli. This way of uplifting vacua guarantees that the interaction of the uplifting sector with the moduli sector is consistent with integrating out the heavy fields in a supersymmetric way. The interactions between  light and heavy fields are characterized in terms of the Kähler invariant function,  $G= K + \log |W|^2$, which is required to be separable in the two sectors. We generalize earlier results that when the heavy fields are stabilized at a minimum of the \emph{Kähler function} $G$ before the uplifting (corresponding to stable AdS \emph{maxima} of the potential), they remain in a perturbatively stable configuration for arbitrarily high values of the cosmological constant (or the Hubble parameter during inflation). By contrast, supersymmetric minima and saddle points of the scalar potential are always destabilized for sufficiently large amount of uplifting.  We prove that  these results remain unchanged after including gauge couplings in the model. We also show that in more general scenarios, where the Kähler function is not  separable in the light and heavy sectors, the minima of the Kähler function still have better stability properties at large uplifting than other types of critical points.
}

\keywords{Flux compactifications, dS vacua in string theory, Supergravity Models}

\preprint{}

\begin{document}
\bibliographystyle{JHEP}

\section{Introduction}
\label{intro}

The search for de Sitter (dS) vacua in string theory has received a lot of attention motivated by the need to construct realistic late-time cosmology scenarios. In 2003 Kallosh, Kachru, Linde and Trivedi (KKLT) \cite{Kachru:2003aw} provided the first example of a mechanism to obtain stable de Sitter vacua in the framework of Type IIB string theory. Their two-step approach was to invoke background fluxes and non-perturbative effects in order to \emph{freeze} the heavy moduli present in the compactification \emph{while preserving supersymmetry}, and then add extra supersymmetry breaking effects in a controlled way, i.e. not interfering with moduli stabilization, so that the anti-de Sitter (AdS) minimum would be uplifted to dS.

In practice, in the KKLT paper and in many sequels that discussed mechanisms of uplifting of the AdS minimum, it is assumed that the complex structure moduli are integrated out before supersymmetry breaking effects are taken into account. The effective field theory left after freezing these fields is assumed to be $\cN=1$ supergravity. In other words, the heavy moduli are integrated out \emph{supersymmetrically} \cite{Binetruy:2004hh} and are assumed to be consistently decoupled --in the sense of \cite{deAlwis:2005tg,deAlwis:2005tf,Achucarro:2008sy}-- from the light fields.

The problem, as discussed in \cite{Achucarro:2008sy,Choi:2004sx,deAlwis:2005tg,deAlwis:2005tf,Achucarro:2007qa,BenDayan:2008dv}, is that in general it cannot be taken for granted that the supersymmetry breaking corrections added to the effective action are consistent with the process of \emph{supersymmetrically} integrating out part of the moduli. The reason is very simple: the heavy fields should be integrated out in the \emph{full  theory}, including the supersymmetry breaking modifications. Any other way to proceed may lead to inconsistencies.  For example, the minimum where the moduli were stabilized might shift by the supersymmetry breaking effects, in which case the low energy effective theory would have the heavy fields --the complex structure moduli in this case-- fixed at a point which is not even an extremum of the scalar potential.

In this article we will study  a mechanism of $F$-term uplifting  \cite{GomezReino:2006dk,GomezReino:2006wv,GomezReino:2007qi} consistent with the supersymmetric integration of the  heavy moduli \cite{Achucarro:2008sy,Achucarro:2007qa}. The basic idea of $F$-term uplifting  consists of adding an extra sector to the theory defined by the Kähler and complex structure moduli which breaks supersymmetry separately,  lifting the vacuum to dS. In order to avoid that the interactions between the two sectors spoil the stabilization of moduli it is required that they are only weakly coupled. In the original papers on $F$-term uplifting this was achieved  by requiring that the two sectors interact only with gravitational strength, i.e. coupling the sectors as: 
\be 
  K = K^{(moduli)} +K^{(uplift)} \qquad W = W^{(moduli)}+W^{(uplift)}, 
  \label{GRansatz}
\ee 
and requesting all dimensionful couplings in the uplifting sector to be small compared to the Planck mass. However   this ansatz does not satisfy in general the  necessary conditions for consistent supersymmetric decoupling  of the heavy moduli \cite{deAlwis:2005tg,deAlwis:2005tf,Achucarro:2008sy,BenDayan:2008dv}. For later developments on F-term uplifting mechanism see \cite{Saltman:2004sn,Lebedev:2006qq,Lebedev:2006qc,Dine:2006ii,Kitano:2006wm,Dudas:2006gr,Kallosh:2006dv,Abe:2006xp,Abe:2007yb,Lalak:2007qd,Brax:2007xq,Abe:2008ka,Papineau:2008xf,Everett:2008qy,Covi:2008ea,Aparicio:2008wh,Everett:2008ey,Blumenhagen:2008kq,deAlwis:2008kt}.

In these types of models it is tempting to argue that the effects of truncating the heavy fields  inconsistently would be too small to affect seriously the physics of the effective theory. Although this might be correct if we are only interested in low energy phenomenology, when the effective theory is used to describe inflation the situation is much more subtle.  In this scenario the inflationary sector is what plays the role of the supersymmetry breaking sector and, as in the case of uplifting, its interactions with the moduli fields have to be consistent with the supersymmetric integration of the heavy moduli.  Recently Davis and Postma \cite{Davis:2008sa} discussed an enlightening example that illustrates the problems of an inconsistent truncation in an inflationary model.  They studied the $F$-term hybrid inflationary model proposed in \cite{Kallosh:2004yh} which includes a moduli sector of the KKLT or racetrack \cite{BlancoPillado:2004ns} form. This model gives viable inflation as long as the volume modulus is assumed to be fixed during inflation and some of the parameters are fine-tuned.  However, this truncation of the modulus field is not consistent. When the dynamics of this field are taken into account it can be seen that the field does not remain constant during the inflationary period, it shifts from its stable value at the end of inflation. The shift results in corrections to the inflationary potential that spoil its flatness and therefore ruin inflation (see e.g. \cite{BenDayan:2008dv} for a recent discussion and references).

In   a recent paper \cite{Achucarro:2008sy} we revisited the conditions for consistent supersymmetric decoupling  of the heavy moduli. We found that these conditions can be translated into a particular form of the Kähler invariant function $G= K + \log |W|^2$. For example, if the Kähler potential is separable, $K = K^{(h)}(heavy) + K^{(l)}(light)$, it is sufficient to require that the full Kähler invariant function is also separable: 
\be 
  G = G^{(h)}(heavy) + G^{(l)}(light), 
  \label{Introansatz} 
\ee 
or, equivalently, that the superpotential factorizes in the two sectors: 
\be 
  W= W^{(h)}(heavy) W^{(l)}(light). 
\ee Note that consistent decoupling does not require the scalar manifold to be a product, and thus this is just a special case of the class of interactions consistent with the supersymmetric integration of the heavy fields.
 
The ansatz (\ref{Introansatz}) has a long history.  In the early 80's it was studied as mechanism  to couple the visible matter fields to a supersymmetry breaking sector or the inflationary  sector \cite{Cremmer:1982vy,Binetruy:1984wy}, and  more recently  has been discussed as a sufficient condition  to  integrate out heavy chiral multiplets in a  supersymmetric way \cite{Binetruy:2004hh}. It has also appeared in connection with brane inflationary models and moduli stabilization, in particular in the $D3/D7$ model \cite{Hsu:2003cy}, where it was shown that the ansatz preserves the AdS flat direction (from shift symmetry) due to the BPS character of the configuration.

In \cite{Achucarro:2007qa} we studied the possibility of using a separable  Kähler function (\ref{Introansatz}) as an alternative way  to couple the heavy moduli to the supersymmetry breaking sector in $F$-term uplifting mechanisms. This type of coupling ensures that if the heavy fields are integrated out at a supersymmetric critical point they remain at a critical point of the potential after adding the supersymmetry breaking sector.  Moreover, the F-terms of the heavy moduli remain zero after the uplifting, and thus supersymmetry breaking receives no contribution from the heavy fields.

It is remarkable that, in spite of the direct couplings present in $W$, the light and heavy sectors almost do not interact \cite{Achucarro:2007qa, Achucarro:2008sy} even when supersymmetry is broken by the light sector.  Actually using the ansatz (\ref{Introansatz}) the perturbative stability analysis of the uplifted vacuum decouples in the two sectors. In particular the stability condition along the heavy field directions has a simple form, it has no dependence on the details of the uplifting sector other than through a single parameter that measures the amount of uplifting, $H / m_{3/2}$, the ratio of the Hubble expansion rate to the gravitino mass of the uplifted vacuum. In \cite{Achucarro:2007qa} we analyzed a toy model with a single heavy field and found a region in parameter space where the critical points of the heavy sector remain stable\footnote{In this paper, as in reference \cite{Achucarro:2007qa}, we only study the perturbative stability of the uplifted vacua and therefore, after the uplifting, by stable vacua we mean \emph{local} minima of the scalar potential. } for arbitrary values of this uplifting parameter $H/m_{3/2}$.  Interestingly, these critical points are \emph{stable} AdS maxima before the uplifting, and in our model correspond to minima of the Kähler function $G^{(h)}(heavy)$. In this paper we will prove that this result can be extended to an arbitrary number of chiral fields in the heavy supersymmetric sector, provided they satisfy (\ref{Introansatz}),  and that it survives the inclusion of gauge interactions. Also, in more general scenarios where the Kähler function is not required to be separable, we will prove using mild assumptions  that the supersymmetric AdS maxima of the potential always become  stable for large enough values of the uplifting parameter. The remarkable stability of dS vacua resulting from highly uplifted AdS local maxima had not been noticed before and constitutes one of the main results in the paper.

Our work complements those of Gómez-Reino and Scrucca \cite{GomezReino:2006dk,GomezReino:2006wv,GomezReino:2007qi} and, more recently, Covi \emph{et al.}  \cite{Covi:2008ea}, who give necessary conditions for the stability of uplifted vacua along the supersymmetry breaking directions, which we include in the light sector. Here instead we obtain necessary and sufficient conditions for the stability of the supersymmetrically integrated moduli, about which we have little information or observational input. We take for granted the existence of a stable dS vacua in the effective theory for the light sector, which therefore has to satisfy the conditions from \cite{Covi:2008ea}.  We will return to this point at the end of the paper.

In this work we will not make any specific assumptions about the origin of the set of fields that are integrated out supersymmetrically. For example in the KKLT framework they could be identified with both Kähler and complex structure moduli, or with the complex structure moduli alone depending on the masses of each set of fields. However it should be clear that our results are not restricted to type IIB compactifications. Regarding the light sector we expect it  to include both the visible sector and the hidden sector where supersymmetry is broken spontaneously.   \\

This paper is organized as follows. We begin in section \ref{RecallResults} with a quick review of the relevant results and notation. In particular we recall the basic features of the $F$-term uplifting mechanism characterized by the ansatz (\ref{Introansatz}). In section \ref{stability1} we study the relation between the critical points of the Kähler invariant function $G$ and the scalar potential and prove a one-to-one correspondence between the minima of $G$ and the supersymmetric (AdS) local maxima of the potential. The results in section \ref{stability1} are completely general -- we make no assumptions about the form of $G$, only that supersymmetry is unbroken. At the same time we introduce the technique used in later sections for the stability analysis of the uplifted vacua. In section \ref{stability2} we analyse the stability of the heavy fields in F-term uplifted vacua where the coupling to the light (supersymmetry breaking) sector is given the ansatz (\ref{Introansatz}).  We extend the results of reference \cite{Achucarro:2007qa} to an arbitrary (supersymmetric) heavy sector; we also consider the effect of gauge couplings and D-terms and show that the results are unchanged for consistently decoupled charged fields. Finally, in section \ref{generalcouplings}, we consider more general uplifting scenarios where the coupling between the light and heavy fields no longer satisfies (\ref{Introansatz}) but only the milder condition $K  = K^{(h)}(heavy) + K^{(l)}(light)$. We finish with a summary of our results and a discussion in section \ref{conclusions}.

\section{$F$-term uplifting and the integration of heavy moduli}
\label{RecallResults}
\subsection{Notation and conventions}
\label{SUGRA}

We work in units of the reduced Planck mass, $8\pi M_p^2 = 1$. We start by recalling that the $\mathcal{N}=1$ supergravity action involving scalars and gauge fields (chiral and gauge superfields) is entirely described by three functions of the scalars: the Kähler potential $K(\xi ,\bar{\xi})$, the holomorphic superpotential $W(\xi)$ and the gauge kinetic functions $f_{ab}(\xi)$. The bosonic part of the action is 
\begin{equation}
  S = \int d^{4}x\sqrt{-g} (\ft{1}{2}R + T + \mathcal{L}_{\mathrm{gauge}} - V ) \label{Sdef}. 
\end{equation}
It can be shown that the action and supersymmetry transformations are invariant under Kähler transformations, 
\begin{equation} 
  \label{kahlertrans} K \to K + h(\xi) + \bar{h}(\bar{\xi}) \qquad W \to W e^{-h(\xi)}, 
\end{equation} with $h(\xi)$
an arbitrary holomorphic function. Actually, if $W \neq 0$, the action can be written in terms of the Kähler invariant function $G = K +\log |W|^2$ and the gauge kinetic functions. In a model with $n_c$ chiral multiplets and $n_v$ vector multiplets, we can write the kinetic terms of the scalar fields $\xi^I$, $\xi^{\bar I} = (\xi^I)^*$ using the Kähler function $G(\xi,\bar \xi)$ as follows 
\be 
  T = G_{I\bar{J}}\; \nabla_{\mu}\xi^{I}\nabla^{\mu} \xi^{\bar J},  \quad \text{where} \quad G_{I\bar{J}} \equiv \partial_{I} \partial_{\bar{J}} G = \partial_{I} \partial_{\bar{J}} K, \qquad I,J= 1,\ldots, n_c. 
  \label{Skinetic} 
  \ee 
Here we have denoted the gauge covariant derivatives by  $\nabla_{\mu}\xi^{I} = \partial_{\mu}\xi^{I} - A_{\mu}^{a}k_{a}^{\phantom{a}I}(\xi)$, and $k_{a}^{\phantom{a}I}(\xi)$ are the Killing vectors that define the gauge transformations of the scalars:
\be 
  \delta_{gauge}\xi^{I}=k_{a}^{\phantom{a}I}(\xi)\alpha^{a},  \qquad  a=1,\ldots,n_v,
\ee
where $\alpha^{a}$ are the gauge parameters. The kinetic terms of the gauge fields are determined by the (holomorphic) gauge kinetic functions $f_{ab}(\xi)$:
\begin{equation} 
\mathcal{L}_{\mathrm{gauge}} = - \frac{1}{4}(\Re f_{ab})F^{a}_{\phantom{a}\mu\nu}F^{b\mu\nu} + \frac{1}{4\sqrt{-g}} (\Im f_{ab})F^{a}_{\phantom{a}\mu\nu}\epsilon^{\mu\nu\rho\sigma}F^{b}_{\phantom{b}\rho\sigma}.
\end{equation}
The scalar potential includes a contribution from F-terms and D-terms 
\begin{equation}
  V =  V_{F} + V_{D},
\end{equation} 
where $V_{F}$ and $V_{D}$ can be written in as a function of the auxiliary fields of the chiral and gauge superfields, $F^{I}$ and $D^{a}$ respectively: 
\begin{subequations} 
  \begin{align} 
    V_{F} &=  G_{I\bar{J}}F^{I}F^{\bar{J}}  - 3e^{G} = \rme^{G} ( G^{I \bar J } G_I G_{\bar J}- 3), \label{Gaction}\\ V_{D} &= \frac{1}{2} \Re(f_{ab}){D}^{a}D^{b} \label{scalard}. 
  \end{align} 
\end{subequations}
The auxiliary fields have equations of motion that can be solved algebraically in terms of the chiral fields: 
\begin{subequations} 
  \begin{align} 
    F^{I} &= e^{G/2}G^{I\bar{J}} {G}_{\bar{J}}\label{fterms} \\ D^{a} &= i(\Re f)^{-1ab}k_{b}^{I}G_{I} = -i(\Re f)^{-1ab}k_{b}^{\bar I}G_{\bar I}. \label{Dterm} 
  \end{align} 
\end{subequations} 
The two expressions given for the D-terms are equivalent due to the  gauge invariance of the Kähler function $G(\xi, \bar \xi)$  \cite{Binetruy:2004hh}:
\begin{equation}
  \delta_{\mathrm{gauge}} G=(k_{a}^{I}G_{I}+k_{a}^{\bar{I}}G_{\bar{I}})\alpha^{a} = 0, \quad  \text{for all} \quad a=1,\ldots,n_v. 
\end{equation}
In this paper we will assume  that there are no constant Fayet-Iliopoulos terms present, since these require a more careful treatment that is outside the scope of our present study.

The $\cN=1$ supersymmetry transformations of the fermions in the chiral and vector multiplets $\chi^I$ and $\lambda^a$ will be of special relevance in the following discussion:
\begin{subequations} 
  \begin{align} 
    \delta \chi^I_{\phantom{I}L} &= \ft12 \gamma^\mu \nabla_\mu \xi^I \epsilon_R - \ft12 \, \rme^{\ft12 K} K^{I \bar J} \cD_{\bar J} \bar W \, \epsilon_L  \\ \delta \lambda^a &= \ft14 \gamma^{\mu\nu} F^a_{\phantom{a}\mu\nu} \epsilon + \ft12 i D^a \gamma_5  \epsilon 
  \end{align} 
  \label{SUSYtrans} 
\end{subequations} 
Here $\epsilon$ is the parameter of the supersymmetry transformations, and $\gamma^\mu$ represent the gamma matrices as usual. The subscripts $R$ and $L$ of the fermions stand for right and left chirality respectively: 
\be
  \chi^I_{\phantom{I}R} = \ft12 (1 - \gamma^5) \chi^I_{\phantom{I}R} \quad \qquad \chi^I_{\phantom{I}L} = \ft12 (1 + \gamma^5) \chi^I_{\phantom{I}L}
\ee

From (\ref{SUSYtrans}) we can see that in a homogeneous background  ($\nabla^{\mu} \xi^I = F^a_{\phantom{a} \mu\nu } = 0$), a set of necessary conditions for unbroken supersymmetry is: 
\be 
  \cD_I W = 0 \qquad \text{for all} \qquad I=1,\ldots, n_c. \label{SUSYcond0}
\ee Equivalently this condition can be written in terms of the Kähler function as: 
\be 
  \partial_I G(\xi,  \bar \xi) =0 \qquad  \text{for all} \qquad  I=1,\ldots, n_c. \label{SUSYcond} 
\ee Note that although it is always possible to break supersymmetry spontaneously by non-vanishing $F$-terms and zero $D$-terms (\ref{SUSYtrans}), the relations  (\ref{fterms}) and (\ref{Dterm}) imply that non-vanishing $D$-terms necessarily  require  non-vanishing $F$-terms, and  therefore supersymmetry can never be broken by $D$-terms alone \cite{Choi:2005ge}.

The result  (\ref{SUSYcond}) implies, together with the expression for the scalar potential (\ref{Gaction}) and (\ref{scalard}), that supersymmetric critical points $\xi^I_0$ with non vanishing superpotential $W(\xi_0)\neq0$  always have a negative vacuum energy, i.e. they are AdS critical points: 
\be
V(\xi_0)  = - 3 e^{G(\xi_0)} <0.
\ee
Interestingly supersymmetric critical points are always perturbatively stable, regardless of being local minima, maxima or saddle points. The reason is that in an  AdS background a fluctuation with a tachyonic mass might still be stable as long as it satisfies the Breitenlohner-Freedman bound \cite{Breitenlohner:1982bm}: 
\be 
  m^2 \ge \frac{3}{4}\; V(\xi_0), \label{BF} 
\ee 
which is always fulfilled by supersymmetric critical points.

\subsection{Uplifting and consistent integration of heavy moduli}
Let us now review the basic features of the $F$-term uplifting mechanism proposed in \cite{Achucarro:2007qa}. In this new class of $F$-term uplifting  the couplings between the  heavy moduli and the uplifting sector are consistent with the requirements found in \cite{Achucarro:2008sy} for the supersymmetric integration of heavy moduli. The coupling between the heavy moduli, $H^\alpha$, and the uplifting sector, which belongs to the set of light fields $L^i$, is defined in terms of an ansatz for the Kähler function of the form (\ref{Introansatz}): 
\be 
  G(H,\bar H,L,\bar L ) = A(H,\bar H) + B(L,\bar L). \label{ansatz} 
\ee 

In the absence of gauge interactions the scalar potential derived from this ansatz can be seen to be, using (\ref{Gaction}): 
\be 
  V = \rme^{A+B} \left( A^{\alpha \bar \beta} A_{\alpha} A_{\bar \beta} +B^{i \bar j} B_{i } B_{\bar j} -3 \right). \label{potential} 
\ee
Fixing the heavy fields at the supersymmetric critical point $H^\alpha_0$ we obtain the effective potential of the low energy theory, which reduces to the simple expression: 
\be 
  V(H_0,L) = \rme^{A(H_0)} V_{\textit{light}}(L), \label{effpot}
\ee 
where $V_{\textit{light}} = \rme^B (B^{i\bar j} B_i B_{\bar j}-3 )$ is  the scalar potential of the uplifting sector when considered alone. The uplifting properties of this ansatz can be summarized as follows:
\begin{itemize}
\item Suppose that $H_0^\alpha$ is a supersymmetric critical point of the heavy sector, and $L_0^i$ is a critical point of $V_{\textit{light}}$, then the field configuration $(H_0^\alpha, L_0^i)$ is a critical point of the full potential.
\item The value of the potential of the light sector $V_{\textit{light}}(L)$ at the critical point  $L^i_0$ determines whether the supersymmetric vacuum  is lifted to dS, Minkowski or remains AdS:
\begin{eqnarray}
V_{\textit{light}}(L)>0 \qquad &\Longrightarrow& \qquad (H_0^\alpha, L_0^i) \text{ is a dS vacuum} \nonumber\\
V_{\textit{light}}(L)=0 \qquad &\Longrightarrow& \qquad (H_0^\alpha, L_0^i) \text{ is a Minkowski vacuum} \nonumber \\
V_{\textit{light}}(L)<0 \qquad &\Longrightarrow& \qquad (H_0^\alpha, L_0^i) \text{ is an AdS vacuum.} \nonumber
\end{eqnarray}
\item If there is more than one supersymmetric configuration of the heavy sector, all of them become degenerate when uplifted to Minkowski (note that this makes the possibility of topological inflation quite natural).
\end{itemize}
In view of the direct couplings in (\ref{potential}) one might think that the two sectors strongly influence each other and therefore the uplifting would easily destabilize the heavy moduli, however in \cite{Achucarro:2007qa} we found that the two sectors almost do not interact. We studied the perturbative stability of vacua of the form $(H_0^\alpha, L_0^i)$ where $H_0^\alpha$ is a supersymmetric configuration of the heavy sector and $L_0^i$ a critical point of $V_{\textit{light}}$. We found that the mass matrix around this vacuum is block diagonal in the two sectors, meaning that there are no quadratic interactions between the fluctuations of the heavy and light fields: 
\be 
  \partial^2_{i \alpha} V (H_0,L_0) =0, \qquad  \partial^2_{i \bar \alpha}  V(H_0,L_0) =0. \label{blockdiag} 
\ee 
Actually it turns out that this  is a prerequisite for freezing the heavy fields consistently. To integrate out the fluctuations with large masses around a given vacuum first we have to find their mass spectrum, which requires diagonalizing the mass matrix, and only after having identified the heavy  modes  can we set them to zero consistently. Proceeding in this way, by construction, the mass matrix at the vacuum is always block diagonal in the massive and light modes.

This result allows us to study the stability of the two sectors separately. In the case of the supersymmetry breaking sector it is possible to prove a remarkably  simple result, namely that the vacuum $(H_0^\alpha,L_0^i)$ is perturbatively stable with respect to fluctuations of the light fields as long as $L_0^i$ is a minimum of the potential of the light sector $V_{\textit{light}}(L)$. For the heavy sector it is more difficult to obtain model independent statements concerning the perturbative stability. Nevertheless  in  \cite{Achucarro:2007qa} we where able to give a few general results: 
\begin{itemize}
\item The stability analysis for fluctuations of the heavy fields depends on the light sector only through a single parameter $b= B^{i \bar j} B_i B_{\bar j}|_{L_0}$ that  controls the amount of uplifting, 
\be 
  b-3 = e^{-G} V = \biggl({ 3 H\over m_{3/2}}\biggr)^2 
\ee
\item Any vacuum becomes stable or neutrally stable with respect of fluctuations of the heavy fields after being uplifted to Minkowski. A similar result was found in \cite{BlancoPillado:2005fn}, where Blanco-Pillado \emph{et al.} argued that SUSY vacua with vanishing cosmological constant are automatically stable, up to flat directions.
\item For large amounts of uplifting the full potential becomes approximately:
\be
  V(H,L) \approx b \, \rme^{A+B},
\ee
and therefore the stable configurations of the heavy sector are those minimizing the Kähler function
$A(H,\bar H)$. 
 \end{itemize}

In order to  understand better the details of this new uplifting mechanism we analyzed the perturbative stability of the supersymmetric sector in a toy model with only one heavy field. The result of this study was quite surprising, we found that the supersymmetric AdS maxima of the potential at zero uplifting ($b=0$), which are stable since they satisfy the Breitenlohner-Freedman bound, remain stable configurations of the heavy sector for any uplifting. Interestingly, we also found that these AdS maxima coincide with the minima of the Kähler function. When we studied the  uplifting of AdS minima of the scalar potential we recovered the standard result, for sufficiently large amount of uplifting these configurations always become unstable.  This result opens a new  door for the construction of stable dS vacua, instead of constraining ourselves to the uplifting of AdS minima we can also use the AdS maxima of the potential which seem to have better stability properties, at least for a certain class of interactions between the moduli and the uplifting sectors.

In \cite{Achucarro:2008sy} we also considered the case where the gauge interactions of the light sector were turned on. We found that, as long as the heavy fields are consistently decoupled, the gauge interactions in the light sector do not change the results listed above. The consistent  decoupling of the heavy sector imposes certain restrictions on the type of allowed gauge interactions. In particular any gauge field that survives at low energies should not interact simultaneously with both heavy and light fields, otherwise the heavy fields could be sourced in the low energy theory due to the gauge interactions. In practice this requirement  means that  the Killing vectors of the light sector can only have components along the light directions, $k^i_a$. Moreover, a consistent decoupling  also demands  that the critical points of the heavy sector do not shift due to the presence of the light sector, implying that the Killing vectors and the gauge kinetic functions of the light sector cannot depend on the heavy fields, $k=k^i_a(L)$, $f_{ab}(L)$. In this situation the contribution to the scalar potential generated by gauge interactions, the D-term potential (\ref{scalard}), is independent of the heavy fields, and thus the stability analysis along the heavy directions is unaffected.

\section{Stability of supersymmetric critical points}
\label{stability1}
In this section we study the stability properties of a supersymmetric critical point in a completely general setup.  We take the action to be characterized by a Kähler potential $G(\xi,\bar \xi)$, and we allow for an arbitrary gauge coupling defined by the gauge kinetic functions $f_{ab}(\xi)$ and Killing vectors $k(\xi)_a^I$.  We will relate the stability of the supersymmetric vacua to the curvature of the Kähler function, and in particular we will show that maxima of the scalar potential always correspond to minima of the Kähler function.

\subsection{Analysis of the Kähler function}
We begin by studying the character of the critical points of the Kähler function, which is a simple calculation and will serve us to introduce the technique  we will  use later in the analysis  of the scalar potential. The Taylor expansion of the Kähler potential $G(\xi^I, \xi^{\bar I})$ around  a supersymmetric critical point  $\xi_0^{I}$, reads:
\be\label{taylorG}
  G= G(\xi_0) + G_I(\xi_0) \hat \xi^I + G_{\bar I} (\xi_0) \hat{\xi}^{\bar I} + G_{I \bar J}(\xi_0)\;  \hat \xi^I \hat \xi^{\bar J} + \ft12 G_{I J}(\xi_0) \; {\hat  \xi}^{I}\hat \xi^{J} + \ft12 G_{\bar I \bar J}(\xi_0) \; {\hat  \xi}^{\bar I}\hat \xi^{\bar J} +\ldots ,
\ee
where we define $\hat \xi^{I} = \xi^{I} - \xi_0^{I}$. Note that  the first order terms vanish since $G_{I}(\xi_0)=0$. In order to know if $\xi_0^{I}$ corresponds to a minimum, a maximum or a saddle point the Kähler function we need to find the eigenvalues of its Hessian evaluated at the critical point
\be
  \begin{pmatrix}
    G_{I\bar J} & G_{I  J} \\
    G_{\bar I \bar J} & G_{\bar I J} 
  \end{pmatrix}_{\xi_0}.
\ee
This expression simplifies considerably by redefining the $\xi^I$ fields so that they have canonical kinetic terms at the critical point, $G_{I \bar J}(\xi_0) = \delta_{I \bar J}$. With this choice of coordinates the equation for the eigenvalues $g$ takes the form:
\be
  \det \begin{pmatrix}
    (1-g)\unity & X \\
    X^\dag &(1-g) \unity
  \end{pmatrix}=0
  \label{Geigenvalues}
\ee
where we have used the matrix notation $X =X^T = G_{I J}(\xi_0)$ and $\unity=\delta_{I \bar J }$. Using a known property of determinants  
\be
  \det \begin{pmatrix}
    M&P\\
    Q&N
  \end{pmatrix} = \det\left(MN -QP \right), \quad \text{provided that\; } QN=NQ \quad \text{and\; } \det(M)\neq0,
  \label{trick}
\ee
for square submatrices $M,N,P,Q$, we can see that $g$ is a solution of (\ref{Geigenvalues}) if and only if it also solves: 
\be
  \det\left((1-g)^2 - X^\dag X\right)=0.
  \label{Geigenvalues2}
\ee 
Strictly speaking this equation was derived for $g \neq1$, but it is not difficult to check that it also gives the correct solution for $g=1$. In order to solve this equation we use the freedom of field redefinition once more. Requiring that the fields have canonical kinetic terms is not enough to fix the choice of fields completely, we can still redefine the fields by a constant unitary transformation of the form $\tilde \xi^{I} = U^I_{\phantom{I }J} \, \xi^{J}$. Under this field redefinition the matrix $X$ and the combination $X^\dag X$ transform as:
\be
  X = U^T \tilde X U, \qquad X^\dag  X = U^\dag \,( \tilde  X^\dag  \tilde X) \,U, \quad \text{where} \quad U=U^I_{\phantom{I}J},
\ee
and therefore we can use this freedom to  transform the Hermitian combination $X^\dag X$ into a real diagonal matrix. The eigenvalues of $X^\dag X$ are necessarily  nonnegative, and we will denote them by $|x_\lambda|^2$, with $\lambda$ labeling the $p$ different eigenspaces. Moreover, the symmetry of $X$ implies that  $X^\dag X = (X X^\dag)^*$, thus in the  basis  that makes $X^\dag X$ diagonal we have:
\be
  X^\dag X = X X^\dag = Diag(|x_1|^2 \unity_{n_1},\ldots,|x_{p}|^2\unity_{n_p}), \qquad |x_\lambda|^2 \ge 0,
  \label{diagform}
\ee
where $n_\lambda$ is the dimension of the eigenspace of eigenvalue $|x_\lambda|^2$. Note also that in  this particular basis the matrices $X$ and $X^\dag X$ commute, which implies that $X$ should be block diagonal in each of the  eigenspaces of $X^\dag X$: 
\be
  X = Diag(X_1, \ldots,X_p) \qquad \text{with} \qquad X^\dag_\lambda X_\lambda = |x_\lambda|^2 \unity_{n_\lambda}.
  \label{Xblocks}
\ee
This means that the eigenvalue problem decouples for the $m$ different eigenspaces of $X^\dag X$, and therefore the equation (\ref{Geigenvalues2}) takes a very simple form:
\be
  \prod_{\lambda=1}^{p} \left[ (1-g_\lambda)^2 -|x_\lambda|^2 \right]^{n_\lambda}=0,
\ee
which we can solve easily giving the eigenvalues
\be
  g_{\pm \lambda} = 1 \pm |x_\lambda|.
  \label{Geigenvalues3}
\ee
which have multiplicity $n_\lambda$. The different possibilities for the character of the critical point $H_0^{\alpha}$ are summarized in the following table: 
\bea
  \text{Local minimum} && |x_\lambda|<1 \quad\text{for all} \; \lambda=1,\ldots,p \nonumber\\
  \text{Saddle point} &&|x_\lambda|>1 \quad \text{for some or all} \; \lambda
  \label{Gstability}
\eea
The result (\ref{Geigenvalues3}) also indicates that, for each eigenvalue of $X^\dag X$ that satisfies $|x_\lambda|^2=1$,  the Kähler function has a flat direction and a local minimum (a trough) along one of the complex directions  $\xi^I$.

\subsection{Analysis of the scalar potential with vanishing $D$-terms}
We will now analyze how the maxima and saddle points of the Kähler function relate to the different types of supersymmetric critical points of the scalar potential. This is especially interesting  because the Kähler function is much easier to study. We will demonstrate a remarkable result: the minima of the Kähler function are in one to one correspondence  with the  supersymmetric AdS maxima of the scalar potential. We start assuming that there are no gauge interactions, and in the next section we will prove this result in full generality. 

The stability analysis of a supersymmetric critical point of the scalar potential can be done using the  same techniques of  the previous subsection.  Consider first its Taylor expansion around  a supersymmetric critical point $\xi_0^I$: 
\be
  V = V(\xi_0) + \ft12 V_{I J}(\xi_0)\; \hat \xi^I \hat \xi^{J}  +\ft12 V_{\bar I \bar J}(\xi_0) \;\hat{  \xi}^{\bar I} \hat{ \xi}^{\bar J}  +  V_{I\bar J}(\xi_0) \; \hat{  \xi}^{ I} \hat{ \xi}^{\bar J} + \ldots  \; ,
\ee
where the  second derivatives of the potential evaluated at the point $\xi^I = \xi_0^I$   can be calculated from (\ref{Gaction}) using that $G_{I}(\xi_0)=0$:
\bea
  V_{I J}(\xi_0) &=& - G_{I J}(\xi_0) \rme^{G(\xi_0)} \\
  V_{I \bar J}(\xi_0) &=& \rme^{G(\xi_0)} \left[ G^{R \bar S} G_{R I} G_{\bar S \bar J}- 2 G_{I \bar J}\right]_{\xi_0}.
  \label{seconder1}
\eea 
In order to determine the character of the critical point $\xi_0^{I}$ we need to find the eigenvalues of the Hessian of the potential, which gives the mass spectrum of the fluctuations around $\xi_0^{I}$. As in the previous subsection we define the fields $\xi^{I}$ so that they have canonical kinetic terms, $G_{I \bar J}(\xi_0)= \delta_{I \bar J}$, and the Hermitian matrix $X^\dag X$ becomes diagonal.  With this choice the Hessian has the simple form:
\be
  \begin{pmatrix}
    V_{I \bar J } & V_{I J}\\
    V_{\bar I \bar J} & V_{\bar I J}
  \end{pmatrix}_{\xi_0} =\rme^{G(\xi_0)} 
  \begin{pmatrix}
    X  X^\dag-2 \, \unity & -X\\
    -X^\dag & X^\dag  X-2 \, \unity
  \end{pmatrix}.
  \label{Hessian1}
\ee
Since in the basis we have chosen $X^\dag X = X X^\dag$ it is easy to check that this matrix also satisfies the first of the conditions necessary to apply (\ref{trick}), and we find that   the equation for the spectrum of masses $m^2$ reads:
\be
  \det\begin{pmatrix}
    (X^{\dag} X  -(2+ e^{-G(\xi_0) }\; m^2  )\,\unity)^2-X^\dag X
  \end{pmatrix} =0.
  \label{spectrum1}
\ee
In order to use (\ref{trick}) we also need to assume that the following matrix is non singular, 
\be
  \det \begin{pmatrix}
     X ^\dag X- (2+e^{-G(\xi_0) }\; m^2 )\, \unity
  \end{pmatrix} \neq 0,
\ee
but after some algebra it is possible to prove that (\ref{spectrum1}) also gives the right result in the singular case. As in the previous section we can use that $X$ has the block diagonal form (\ref{Xblocks})  to show that the eigenvalue problem can be decomposed in each of the eigenspaces of $X^\dag X$. Using this fact the eigenvalue equation (\ref{spectrum1}) can be written  as
\be
  \prod_{\lambda=1}^p \left[ (|x_\lambda|^2 - 2-e^{-G(\xi_0) }\; m^2 )^2- |x_\lambda|^2 \right]^{n_\lambda} =0.
\ee
Therefore the spectrum  of masses at the supersymmetric  critical point is given by:
\be
  m_{\pm \lambda}^2 = \rme^{G(\xi_0)}(|x_\lambda|^2-2 \pm |x_\lambda|) = \rme^{G(\xi_0)}\left(\left(|x_\lambda|^2\pm \ft{1}{2}\right)^2 - \ft{9}{4} \right).
  \label{spectrum2}
\ee 
Each of these energy levels contains $n_\lambda$ different excitations with the same mass. The set of quantities $|x_\lambda|$ determine which type of extremum the supersymmetric critical point $\xi_0^I$ is:
\bea
  |x_\lambda|>2 \quad \text{for all $\lambda$} &\Rightarrow& \text{local AdS minimum,} \nonumber\\
  |x_\lambda|<1 \quad \text{for all $\lambda$} &\Rightarrow& \text{local AdS maximum,} 
  \label{Vstability}
\eea 
and any other combination corresponds to AdS saddle points ($|x_{\lambda}|=1,2$ give flat directions). The result (\ref{spectrum2}) also provides a proof of the stability of all supersymmetric critical points, regardless of the possible negative curvature of the potential. Since all these critical points are AdS, the perturbative stability is determined by the Breitenlohner-Freedman bound (\ref{BF}), which is always satisfied  as is clear from (\ref{spectrum2}):
\be
  m^{2} \geq -\frac{9}{4} e^{G(\xi_{0})} = \frac{3}{4}\; V(\xi_0),
  \label{BF2}
\ee

Now we already have at hand all the results we need to check the claim we made at the beginning of this subsection. Comparing equations (\ref{Gstability}) and (\ref{Vstability}) we see immediately that the supersymmetric AdS maxima of the potential always coincide with the minima of the Kähler function.

\subsection{Analysis of the scalar potential with non-vanishing $D$-terms}
Let us now generalize the result of the previous subsection to the case where the gauge couplings are turned on. In this case we have to add to the scalar potential the contribution from $D$-terms (\ref{Dterm}).  In order to calculate the new contributions to the Hessian of the scalar potential around the critical point $\xi^I_0$ we must find the derivatives of the $D$-term potential at this point. Using that $G_I(\xi_0)=0$ we find:
\bea
  V_{D| IJ} (\xi_0) &=& \ft12 (Re f(\xi_0))^{-1\,  ab} \; k^R_a(\xi_0) k_b^{\bar S}(\bar \xi_0)\;  [ G_{IR} G_{J \bar S} + G_{JR} G_{I \bar S} ]_{\xi_0}, \nonumber \\
  V_{D|I \bar J} (\xi_0) &=& \ft12 (Re f(\xi_0))^{-1\, ab} \; k^R_a(\xi_0) k_b^{\bar S}(\bar \xi_0)\; [ G_{IR} G_{\bar J \bar S} + G_{\bar J R} G_{I \bar S} ]_{\xi_0}
\eea 
As we have done previously we will define the scalar fields $\xi^I$ so that they have trivial kinetic terms $G_{I \bar J} = \delta_{I \bar J}$ and the Hermitian matrix $X^\dag X$ is diagonal. In the case of the $D$-term potential we can simplify the calculations even further making use of the freedom we have to define the gauge fields $A_\mu^a$.  Actually, the action is invariant under constant linear transformations of the gauge fields ${A}_{\mu}^{b} = O^{b}_{a} \, \tilde{A}^{a}_\mu$, with $O_{a}^{b}$ any non-singular real matrix, provided that the gauge kinetic functions $f_{ab}$ and the Killing vectors $k^I_a$ transform as follows: 
\be
  \tilde f_{cd} = O^{a}_c \, O^{b}_d \; f_{ab} \qquad \tilde k_b^I = O^a_b \; k_a^I. 
\ee 
In particular note that the gauge covariant derivatives and the Yang-Mills terms do not transform under these redefinitions since:
\be
 A_\mu^a k^I_a  =  \tilde A_\mu^b \tilde k^I_b, \qquad (\mathrm{Re} f_{ab}) F^{a}_{\mu \nu} F^{a \, \mu \nu}  =  (\mathrm{Re} \tilde f_{cd}) \tilde F^{c}_{\mu \nu} \tilde F^{d \, \mu \nu}. 
\ee 
Then we can always use this freedom to turn $\Re f_{ab}$ into a matrix proportional to the identity
\be 
  \Re f_{ab} = e^{G(\xi_0)} \; \delta_{ab}, 
\ee 
where the overall factor has been chosen for convenience. Using these conventions, and defining the matrix $k = k_a^I(\xi_0)$, we can write the Hessian of the $D$-term potential as: 
\be
  \begin{pmatrix}
    V_{D | I \bar J } & V_{D|I J}\\
    V_{D|\bar I \bar J} & V_{D | \bar I J}
  \end{pmatrix}_{\xi_0} =\ft12 \rme^{G(\xi_0)}
  \begin{pmatrix}
    kk^\dag  + X \, kk^\dag X^\dag &  X \,kk^\dag +k^*k^T \, X\\
    X^\dag \, k^*k^T + kk^\dag \, X^\dag & k^*k^T  + X^\dag \,k^*k^T\, X
  \end{pmatrix},
  \label{gaugeHessian}
\ee
which has to be added to (\ref{Hessian1}) in order to  get the Hessian of the full scalar potential.

Before we continue  the calculation let us derive  a  useful property of the Killing vectors $k$. We mentioned in the section (\ref{SUGRA}) that the Kähler  function $G(\xi^I)$ has to be invariant under gauge transformations. In particular in a Taylor expansion of the Kähler function around $\xi_0$ (\ref{taylorG}) all the terms have to be invariant under gauge transformations order by order in $\hat \xi = \xi- \xi_0$. From the gauge transformation of the order one terms in the expansion we find the condition:
\be
  \Big(G_{IJ}(\xi_0) \, k^J_a(\xi_0)  + G_{I \bar J}(\xi_0) \, k^{\bar J}_a(\xi_0) + G_I(\xi_0) \, \partial_J k^I_a \Big) \; \hat \xi^I \alpha^a=0, 
\ee
which has to be satisfied for all values of the gauge parameters $\alpha^a$, and the fluctuations $\hat \xi^I$. Since $G_I(\xi_0) =0$, then with our field definitions and in matrix notation this condition reads simply:
\be
  k^* = - X k.
  \label{projector}
\ee
An immediate consequence of this requirement  is  that the Killing vectors are eigenvectors of the matrix $X^\dag X$ with eigenvalue $g_\lambda=1$:
\be
  X^\dag X \, k =  - X^\dag k^* =   k, 
  \label{eigenk}
\ee
since $X^T = X$. This means that the matrices $kk^\dag$ and $k k^T$ have all the entries zero except in the block that corresponds to the eigenspace of eigenvalue $|x_\lambda|^2=1$ of $X^\dag X$.  As we saw in the previous section the eigenvalue problem of the Hessian decouples in the different eigenspaces of the matrix $X^\dag X$. We have just proven that the corrections introduced by the $D$-term potential respect this decoupling and moreover, that the corrections only affect the eigenspace with eigenvalue $|x_\lambda|^2=1$. Therefore we can use the results of the previous section for all the eigenspaces with $|x_\lambda|\neq1$ to find the corresponding eigenvalues. In the remaining of this section we will just focus on solving the eigenvalue problem in the eigenspace where $|x_\lambda|^2=1$, which we label by $\lambda=1$. In order to keep notation simple, we will use the matrices $X_1$ and $k_1$ to represent the submatrices corresponding to the eigenspace $\lambda=1$, thus: 
\be
  X_1^\dag X_1 = X_1 X^\dag_1 =\unity_{n_1}.
  \label{Xunity}
\ee
Notice that, since the hermitian matrix $k_1 k^\dag_1$ transforms under scalar field redefinitions in the same way as $X_1 X_1^\dag$,   we can use the residual freedom to choose the eigenvectors in the eigenspace with $|x_\lambda|^2=1$ to turn $k_1 k^\dag_1$ into  a real diagonal matrix:
\be
  k_1 k_1^\dag = Diag(|k_1|^2,\ldots,|k_{n_1}|^2),
  \label{diagk}
\ee
where $n_1$ is the dimension of the eigenspace with $|x_\lambda|^2=1$. The final expression for the Hessian of the full potential restricted to this eigenspace can be obtained from (\ref{gaugeHessian}) and (\ref{Hessian1})
\be
  \begin{pmatrix}
    V_{I \bar J } & V_{I J}\\
    V_{\bar I \bar J} & V_{\bar I J}
  \end{pmatrix}_{\lambda =1} = \rme^{G(\xi_0)}
  \begin{pmatrix}
    -\unity  + k_1k^\dag_1  &   (-\unity + k_1k_1^\dag )\,  X_1\\
    (-\unity  + k_1k_1^\dag) \; X_1^\dag & -\unity  + k_1k_1^\dag
  \end{pmatrix},
  \label{totalHessian}
\ee
where we have used the properties (\ref{projector}), (\ref{diagk}) and (\ref{Xunity}) in order to simplify (\ref{gaugeHessian}). It is easy to check that the matrices $X_1$ and $k_1 k_1^\dag$ commute, therefore we can use equation (\ref{trick}) in order to find the equation for the mass spectrum $m^2$, which reads:
\be
  \prod_{i=1}^{n_{1}} \left( \left( |k_i|^2 -1-e^{-G(\xi_0)} \; m^2 \right)^2 - (|k_i|^2 -1)^2 \right) =0,
\ee
after having substituted the diagonal form of $k_1k_1^\dag$ (\ref{diagk}). The solutions to this equations, together with the results we found in the previous section, which apply for $|x_\lambda|\neq1$ are summarized below:
\bea
  m^2_{\pm \lambda} &=& e^{G(\xi_0)}\left(\left(  |x_\lambda \pm \ft12|\right)^2 - \ft94\right)  \qquad \text{if $|x_\lambda|^2\neq1$,}\nonumber \\
  m^2_{+1 i}  &=& 2 e^{G(\xi_0)} \left( |k_i|^2 - 1 \right)  \qquad \; \;  \qquad \text{if $|x_\lambda|^2=1$,} \nonumber \\
  m^2_{-1 i}  &=& 0\phantom{\, e^{G(\xi_0)} \left( |k_i|^2 - 1 \right)  \qquad  \qquad \; \; } \text{if $|x_\lambda|^2=1$.}
  \label{gaugespectrum}
\eea
The quantities $|k_i|^2$ determine the mass of the gauge bosons at the supersymmetric critical point $\xi_0$, therefore we can see that the breaking of gauge symmetries  can only improve the stability of vacuum. This result agrees with the analysis of Gomez-Reino and Scrucca of the stability of uplifted vacua in \cite{GomezReino:2007qi}.

The fact that the Killing vectors are associated to the eigenvalues $|x_\lambda|^2=1$ should not be surprising.  On the one hand the eigenvalues $|x_\lambda|^2=1$ are always related to marginally stable directions $m^2=0$. And on the other hand we know that the potential has to be invariant under gauge transformations, thus each Killing vector has to be naturally associated with a flat direction of the potential, which appear in the spectrum as massless fluctuations (the would-be Goldstone bosons that disappear due to the Higgs mechanism).

In view of the result (\ref{gaugespectrum}) we can argue that the presence of non-vanishing gauge couplings does not modify the conclusion of the previous section, the minima of the Kähler function $G(\xi^I, \xi^{\bar I})$ are always in one to one correspondence with the supersymmetric AdS maxima of the scalar potential.

\section{Stability of uplifted vacua}
\label{stability2}
We now return to the main question in this paper, the perturbative stability of the heavy sector when these AdS supersymmetric vacua are uplifted to dS by supersymmetry breaking effects in the light sector. In this section we again assume that the Kähler invariant function is separable in the heavy and light sectors, eq.  (\ref{ansatz}).

\subsection{Stability of uplifted vacua with zero $D$-term potential}
We start by generalizing the results of \cite{Achucarro:2007qa} to an arbitrary number of heavy fields in the supersymmetric sector.  We assume all fields are uncharged.  

Suppose that the set of chiral fields $\xi^I$ can be split in two sectors: $n_h$ heavy fields $H^\alpha$ and $n_l$ light fields $L^i$, which are coupled as in (\ref{ansatz}). We will assume that the system is stabilized at a critical point of the potential $(H_0^\alpha,L^i_0)$, which is also a supersymmetric configuration of the heavy sector, $G_\alpha(H_0,L_0)=0$, but supersymmetry is broken in the light sector, $G_i(H_0,L_0) \neq 0$.  Then as discussed in section \ref{RecallResults}, the Hessian of the full potential has a block diagonal form in the two sectors (\ref{blockdiag}), and therefore it is consistent to consider the stability of the potential only along the ``heavy'' and ``light'' directions independently. We will take the light sector fixed at a perturbatively stable configuration, and we will focus on the stability analysis along the heavy directions. For this purpose we only need to calculate $V_{\alpha \bar \beta}(H_0,L_0)$ and $V_{\alpha \beta}(H_0,L_0)$ from (\ref{potential}): 
\bea
  V_{\alpha \bar \beta}(H_0,L_0) &=& \rme^{A+B}|_{H_0,L_0} \left[A^{\gamma \bar \delta} A_{\alpha \gamma}A_{\bar \beta \bar \delta} + (b-2) A_{\alpha \bar \beta}\right]_{H_0} \nonumber\\
  V_{\alpha \beta}(H_0,L_0) &=& \rme^{A+B}|_{H_0,L_0} \; (b-1) A_{\alpha \beta}(H_0),
  \label{seconder2}
\eea
where we are using the notation $b=B^{i \bar j} B_i B_{\bar j}|_{L_0}$. We will use same choice of fields as in the previous section, where $H^{\alpha}$ are canonically normalized at $H_0^{\alpha}$  and the matrix $X^\dag X$ is real and diagonal, with $X=A_{\alpha \beta}(H_0)$. With this choice we obtain the following expression for the Hessian of the potential at $(H_0^{\alpha},L_0^i)$:
\be
  \begin{pmatrix}
    V_{\alpha \bar \beta } & V_{\alpha \beta}\\
    V_{\bar \alpha \bar \beta} & V_{\alpha \bar \beta}
  \end{pmatrix}_{H_0,L_0} =
  \rme^{A+B}|_{H_0,L_0} \begin{pmatrix}
    X  X^\dag+(b-2) \, \unity & (b-1)\,X\\
    (b-1)\, X^\dag & X^\dag  X+(b-2) \, \unity
  \end{pmatrix}.
  \label{Hessian2}
\ee

\begin{figure}
 \centering \includegraphics[width=0.4\textwidth]{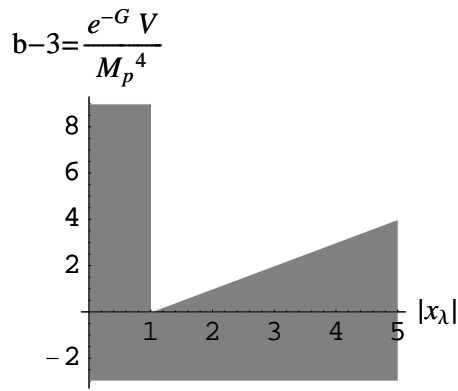}
 \caption{
   Stability of supersymmetric critical points after the uplifting. The quantity on the  vertical axis $b-3$ is  proportional to the cosmological constant (or the Hubble parameter during inflation). The horizontal axis represents the curvature of the Kähler function at the critical point along one of the heavy field directions $H^\alpha$: $|x_\lambda|<1$ corresponds to local minima and  $|x_\lambda|>$ to saddle points.  The shaded region represents stable configurations under perturbations of the heavy fields. For $b<3$ and $b=3$ the uplifted vacua, which are AdS and  Minkowski respectively, are always stable. Local AdS minima of the scalar potential at zero uplifting, $|x_\lambda|>2$, are always destabilized for large uplifting. Local AdS maxima, $|x_\lambda|<1$, remain stable for arbitrary large uplifting.
 }
 \label{fig1}
\end{figure}

Calculating the mass spectrum as in the previous section we arrive at our final result: 
\be 
  m^2_{\pm\lambda} =\rme^{A+B}|_{H_0,L_0} \Big[|x_\lambda|^2 +(b-2) \pm|(b-1) x_\lambda| \Big] =\rme^{A+B}|_{H_0,L_0}\left[ \Big( |x_{\lambda}|\pm\ft12 (b-1)\Big)^2-\ft14(b-3)^2\right]. 
\ee 
To obtain the last expression we assumed that $b>1$, but in the case $b<1$ then $m_{+\lambda}^2$ and $m^2_{-\lambda}$ have to be exchanged. For each energy level characterized by $ m^2_{\pm\lambda}$ there are $n_\lambda$ different excitations with the same mass, where, in analogy with the previous sections,  $n_\lambda$ represents the dimension of the eigenspace of $X^\dag X$ with eigenvalue $|x_\lambda|^2$. The stability condition after uplifting the minimum of the potential to Minkowski or de Sitter, $b\ge3$, reduces to $m^2_{\pm \lambda}>0$ for all $\lambda=1,\ldots,p$, but if the minimum remains AdS after the uplifting, $b<3$, the masses have to satisfy the Breitenlohner-Freedman bound (\ref{BF}):
\bea \label{massmatrixev}
  \text{for } \;b<3\quad  & \Longrightarrow &\quad \left[ \Big( |x_{\lambda}|\pm\ft12 (b-1)\Big)^2-\ft14(b-3)^2\right]\ge \frac{3}{4} (b-3), \nonumber\\
  \text{for } \; b\ge3 \quad & \Longrightarrow & \quad \left[ \Big(|x_\lambda|\pm\ft12 (b-1)\Big)^2-\ft14(b-3)^2\right]\ge 0.  
\eea
Recalling that $b\ge0$, and after a little bit of algebra, it is possible to see that the first of the two inequalities is always satisfied. The second shows that there are no instabilities when the minimum is uplifted to Minkowski $b=3$, although zero modes are possible if any $|x_\lambda|=1$. Thus, the instabilities can only arise for upliftings to dS. These results are summarized in figure \ref{fig1}.

We can see that all the results we obtained in the study of a single modulus toy model in \cite{Achucarro:2007qa} can be generalized to an arbitrary number of fields in the heavy sector. Local AdS minima and saddle points before the uplifting are only stable for small values of the cosmological constant, while local AdS maxima of the potential, which coincide with the local minima of the Kähler function, are always stable.

\subsection{Stability of uplifted vacua with a non-zero $D$-term potential}
\label{stabilitygauge}
Now we study the stability of uplifted vacua when the gauge couplings are turned on.  Including gauge interactions is especially relevant in the case of the light sector, since it includes the visible sector. In section \ref{RecallResults} we mentioned that any gauge field that appears in the effective theory cannot be coupled to the fields that are integrated out, otherwise the gauge fields could act as sources for the truncated fields leading to an inconsistency. For instance, if the truncated fields acquire an expectation value, any gauge field coupled to it would be Higgsed and the full massive vector multiplet would have to be integrated out as well. In our analysis, we will assume that Killing vectors and the gauge kinetic functions satisfy the requirements for a consistent decoupling of the heavy sector  described in section \ref{RecallResults}. Moreover, we will ask the gauge fields in the heavy and light sectors to have decoupled kinetic terms, or in other words, that the gauge kinetic function $f_{ab}$ is block diagonal in the two sectors.  Other than that we will allow the Killing vectors and the gauge kinetic functions of the heavy sector to have an arbitrary dependence on the light and heavy fields $k^{\alpha}_a(H, L)$ and $f_{ab}^{(h)}(H,L)$.  Under these requirements the $D$-term potential reads
\bea V_D
  &=&\ft12\big( Re f(L) \big)^{-1 \, ab} \; k_a^i(L) \, k_b^{\bar j}(\bar L) \; G_i G_{\bar j} + \nonumber \\
  && \ft12\big( Re f(H, L) \big)^{-1 \, ab} \; k_a^\alpha(H, L) \, k_b^{\bar \beta}(\bar H, \bar L) \; G_\alpha G_{\bar \beta}.
  \label{VDansatz}
\eea 
Since the gauge kinetic functions and the Killing vectors of the light sector depend only on the light fields, and $G_i(H, L) = A_i(L)$, all the dependence of $V_D$ on the heavy fields comes from the second term in (\ref{VDansatz}). Therefore, using that $G_\alpha(H_0)=0$, it is easy to check that, despite the dependence of the $D$-term potential of the heavy sector on the light fields, the critical points of the heavy sector are preserved. Moreover, the stability analysis along the heavy field directions is also unaffected by the light sector.  For example, the Hessian of the potential at the critical point remains block diagonal in the two sectors, $V_{D|\alpha i} (H_0, L_0)= V_{D|\alpha \bar i} (H_0, L_0)=0$, and the second derivatives of the $D$-term potential along the heavy directions are given by: 
\bea 
  V_{D|\alpha \beta} (H_0, L_0) &=& \ft12\big( Re f(H_0, L_0) \big)^{-1 \, ab} \; k_a^\gamma(H_0, L_0) \, k_b^{\bar \delta}(\bar H_0, \bar L_0) \; G_{\gamma \alpha}  G_{\bar \delta \beta}, \nonumber \\
  V_{D|\alpha \bar \beta}(H_0, L_0) &=& \ft12\big( Re f(H_0, L_0) \big)^{-1 \, ab} \; k_a^\gamma(H_0, L_0) \, k_b^{\bar \delta}(\bar H_0,\bar L_0) \; G_{\gamma \alpha} G_{\bar \delta \bar \beta}. 
\eea

Thus, in order to find the mass matrix of the heavy fields at the critical point $(H^\alpha_0,L_0^i)$, we only have to add the Hessian of the $D$-term potential with respect to the heavy fields to the result we found for the $F$-term potential (\ref{Hessian2}).  We choose our heavy scalar fields so that they have trivial kinetic terms $G_{\alpha \bar \beta}=1$ and that the matrices $X^\dag X$ ($X = G_{\alpha \beta}$) and $k^\alpha_a k^{\bar \beta}_a$ are both real and diagonal. We also define the gauge fields $A^{(h)\, a}_\mu$ such that the real parts of the gauge kinetic functions of the heavy sector are proportional to the identity matrix:
\be
  \Re f_{ab}^{(h)}  = \rme^{G}|_{H_0,L_0} \, \delta_{ab}.
\ee
The calculation of the $D$-term contribution to the Hessian can be done along the same lines as in section (\ref{stabilitygauge}). Is not difficult to check that the properties (\ref{projector}) and (\ref{eigenk}) still hold, thus here again the matrices $k k^\dag$ and $kk^T$, with $k=k^\alpha_a$, have non-vanishing components only in the eigenspace corresponding to the eigenvalue $|x_\lambda|^2=1$. Thus, after some simplifications, the Hessian of the total scalar potential reads:
\be
  \begin{pmatrix}
    V_{\alpha \bar \beta } & V_{\alpha \beta}\\
    V_{\bar \alpha \bar \beta} & V_{\alpha \bar \beta}
  \end{pmatrix}_{H_0,L_0} =
  \rme^{A+B}|_{H_0,L_0} \begin{pmatrix}
    X  X^\dag+kk^\dag +(b-2) \, \unity & (b-1 +k k^\dag )\,X\\
    (b-1+ kk^\dag)\, X^\dag & X^\dag  X+ kk^\dag + (b-2) \, \unity
  \end{pmatrix}.
  \label{Hessian3}
\ee
From this expression it is straightforward to find the mass spectrum of fluctuations of the heavy sector along the heavy directions:
\bea
  m^2_{\pm \lambda} &=& e^{G(\xi_0)}\left(\left(  |x_\lambda \pm \ft12(b-1)|\right)^2 - \ft14(b-3)^2\right)  \qquad \text{if $|x_\lambda|^2\neq1$,}\nonumber \\
  m^2_{+ 1 i}  &=& 2 \; e^{G(\xi_0)} \left( |k_i|^2 +b - 1 \right)   \quad \qquad \qquad\qquad \qquad \text{if $|x_\lambda|^2=1$,} \nonumber \\
  m^2_{- 1 i}  &=& 0\phantom{\, e^{G(\xi_0)} \left( |k_i|^2 - 1 \right)  \quad \qquad \qquad \qquad\qquad \qquad } \text{if $|x_\lambda|^2=1$.}
  \label{UpliftGaugeMass}
\eea
We can see that  figure \ref{fig1} is still valid when we include the gauge interactions. The only difference with the result in the previous section is that if some of the gauge symmetries are spontaneously broken the mass degeneracy in the eigenspace with $|x_\lambda|=1$  is destroyed. From (\ref{UpliftGaugeMass}) we can see that the presence of gauge interactions only increases the stability of the critical point. 

\section{More general couplings}
\label{generalcouplings}
An interesting question to consider is whether it is possible apply our results to other systems where light and heavy moduli are not coupled with the ansatz (\ref{ansatz}). Let us assume only the mild condition that  the Kähler potentials are separable in the light and heavy sectors: 
\be
  K = K^{h}(H, \bar H) + K^{l}(L,\bar L) \nonumber.
\ee
This condition ensures that mixed derivatives of the Kähler function of the form $G_{i \bar \alpha}(H_0,L_0)$, $G_{i \bar \alpha \beta}(H_0,L_0)$ et cetera...   involving both holomorphic and antiholomorphic indices from the two sectors must vanish. We will also focus on cases where supersymmetry is unbroken at low energies, thus we take the heavy fields fixed at a supersymmetric critical point.

As we discussed in section \ref{RecallResults},  the condition that the potential  has to be block diagonal on the light and heavy fields is necessary  in order to integrate out the heavy fields consistently. Therefore, in any scenario where part of the moduli are going to be integrated out the stability of these fields can be studied independently considering only the "heavy" directions in field space. In order to recover a mass matrix of the form (\ref{Hessian2}) and (\ref{Hessian3}) we would need to satisfy the  extra condition:
\be
  G_{i \alpha}(H_0,L_0)=0.
  \label{condition}
\ee
We can prove this equation from the requirement that the low energy effective action must be invariant under supersymmetry transformations. In particular, this requirement means that the supersymmetry transformations cannot generate the fields that we have truncated. Consider  the supersymmetry transformation of the fermions on the heavy sector, which in a homogeneous bosonic background are simply
\be
  \delta \chi^\alpha_{\phantom{\alpha}L}  = - \ft12 \rme^{\frac{K}{2}} \bar W G^{\alpha \bar \beta} G_{\bar \beta}\; \epsilon_L .
\ee
Expanding this expression to first order in the fluctuations of the light fields around the critical point $L^i = L_0^i + \hat L^i$ gives:
\be
  \delta \chi^\alpha_{\phantom{\alpha}L}  = - \ft12 \rme^{\frac{K}{2}} \bar W G^{\alpha \bar \beta} G_{\bar i\bar \beta}(H_0,L_0)\;  \hat L^{\bar i} \epsilon_L .
\ee
We can see that, unless the quantity $G_{i \alpha}(H_0,L_0)$ vanishes, the supersymmetry transformations will generate the fermions of the heavy sector ($W\neq0$). The condition (\ref{condition}) also ensures that the point where we are fixing the heavy moduli is an extremum of the scalar potential:
\be
  V_\alpha(H_0,L_0) = \Big[\rme^G G^{i\bar j} G_{i\alpha} G_{\bar j}\Big]_{H_0,L_0}=0,
\ee
where we have already used that the Kähler potential is separable and $G_{\alpha}(H_0,L_0)=0$.

Since all the mixed derivatives of the Kähler potential  $G_{i \alpha}(H_0,L_0)$ and $G_{i \bar \alpha}(H_0,L_0)$ vanish, it makes sense to study the curvature of $G(H,L)$ at the  critical point $H^\alpha_0$ only along the heavy directions. Thus, we can repeat the analysis of section 3.1 arriving at similar conclusions:
\begin{itemize}
\item The Kähler function $G(L,H)$ has a local minimum at $H_0^\alpha$ along the heavy directions if the eigenvalues of the matrix $X^\dag X$ satisfy the conditions $|x_\lambda| <1$ for all $\lambda=1,\ldots,p$, with $X = G_{\alpha \beta}(H_0,L_0)$. 
\item If any of the eigenvalues of $X^\dag X$ satisfies $|x_\lambda|>1$ the function $G(H,L)$ has a saddle point at $H_0$.
\item  For each eigenvalue of $X^\dag X$ satisfying $|x_\lambda|=1$ the Kähler function has a neutrally stable direction and a minimum along some complex direction  $H^\alpha$.
\end{itemize}
Using all these results, we can now study the stability of the scalar potential along the heavy directions  as in section 3.2. The second derivatives of the scalar potential  are given by
\bea
  V_{\alpha \beta} (H_0,L_0) &=& \rme^G|_{H_0,L_0}\Big[ (b-1) G_{\alpha \beta} + G^{i \bar j} G_{i \alpha \beta} G_{\bar j}\Big]_{H_0,L_0}, \label{seconder31}\\
  V_{\alpha \bar \beta}(H_0,L_0) &=&  \rme^{G}|_{H_0,L_0} \Big[ G^{\gamma \bar \delta} G_{\alpha \gamma} G_{\bar \beta \bar \delta} + (b-2)G_{\alpha \bar \beta} \Big]_{H_0,L_0},
  \label{seconder32}
\eea
where we have used the notation $b=G^{i \bar j} G_i G_{\bar j}|_{H_0,L_0}$.

Note that, apart from the second term in the equation (\ref{seconder31}), the result we have obtained is of the same form as (\ref{seconder2}). If the quantity $G_{i\alpha \beta}$ stays of order $\cO(1)$, the extra term that we have obtained is roughly of order $\cO(b^{1/2})$, which means that for large values of the uplifting, $b\gg3$, it will become subdominant. Therefore, in this limit, the mass matrix becomes proportional to the Hessian of the Kähler function at $H_0^\alpha$:
\be
  \begin{pmatrix}
    V_{\alpha \bar \beta } & V_{\alpha \beta}\\
    V_{\bar \alpha \bar \beta} & V_{\alpha \bar \beta}
  \end{pmatrix}_{H_0,L_0} =
  b  \begin{pmatrix}
    \unity & X\\
    X^\dag &  \unity
  \end{pmatrix}\rme^{A+B}|_{H_0,L_0} = b 
  \begin{pmatrix}
    G_{\alpha \bar \beta} & G_{\alpha \beta}\\
    G_{\bar \alpha \bar \beta}&  G_{\bar \alpha \beta}
  \end{pmatrix}_{H_0,L_0} \rme^{A+B}|_{H_0,L_0},
\ee 
indicating that the minima of the Kähler function along the heavy directions will always survive uplifting to an arbitrary large value of the cosmological constant. Note also that before uplifing, $G_i(H_0,L_0)=0$, the mass matrix given by (\ref{seconder32}) coincides with (\ref{seconder1}), so we can again identify the AdS maxima of the scalar potential with the local minima of the Käher function along the heavy directions.

We would like to emphasize that in order to obtain this result we have made very mild assumptions. We have required that the  Kähler  potential is separable in the two sectors, we have also imposed the condition that the effective action left after integrating out the heavy moduli is invariant under supersymmetry, and finally we asked the quantity $G_{i \alpha \beta}$  to stay of order $\cO(1)$ for large values of the uplifting. In this scenario we have proved that the AdS maxima of the potential along the heavy directions at zero uplifting ($G_i(H_0,L_0)=0$), which  are perturbatively  stable configurations, remain stable after the uplifting for arbitrary large values of the cosmological constant.

\section{Summary and Conclusions}
\label{conclusions}
In this paper we have studied in detail the stability properties of the $F$-term uplifting mechanism recently proposed in \cite{Achucarro:2007qa}. This way of uplifting AdS vacua guarantees that the interactions between the uplifting sector and the moduli of the compactification are consistent with  integrating out the heavy fields in a supersymmetric way  \cite{Achucarro:2008sy}.  The exact composition of the heavy sector in a KKLT scenario depends on the details of the compactification, but we expect it to include the complex structure moduli and some heavy Kähler moduli. In that  case the light sector would comprise the remaining  Kähler moduli, the visible matter fields and the hidden sector where supersymmetry is spontaneously broken.

In this type of $F$-term uplifting mechanisms the couplings between  the light fields $L$ and heavy fields $H$ are characterized by the separability of the Kähler invariant function of the total theory
\be
  G(H,\bar H,L,\bar L)=G^{(h)} (H,\bar H) + G^{(l)}(L,\bar L), \nonumber 
\ee
which can be expressed in terms of the Kähler potential and the superpotential as follows:
\bea
  K(H,\bar H,L,\bar L) &=& K^{(h)}(H,\bar H)  + K^{l}(L,\bar L) \nonumber\\
  W(H,L) &=& W^{(h)}(H)\, W^{(l)}(L)\nonumber.
\eea
This ansatz is approximately satisfied by the couplings between the frozen complex structure moduli and the Kähler moduli in large volume scenarios \cite{Conlon:2005ki,Conlon:2006wz,Conlon:2007dw}. In these models it ensures the consistency of including the non-perturbative effects  with the supersymmetric integration of the complex structure moduli\footnote{We thank Joe Conlon for a discussion on this point.}. 

The key property of this type of coupling is that  the heavy fields remain at a supersymmetric configuration after coupling them to the light sector, even when supersymmetry is broken by the light fields. In view of the direct couplings in the superpotential it might appear that the two sectors are strongly interacting, and thus that the  supersymmetry breaking is likely to spoil the stabilization of the heavy moduli. However, a more careful analysis reveals that the two sectors are essentially decoupled. For instance, the mass matrix of field fluctuations around the uplifted vacuum is block diagonal in the light and heavy directions. This implies that the stability analysis can be done independently for the light and heavy sectors. In this paper we have focused on the study of the stability of the heavy moduli that are integrated out. Our results completely confirm and generalize those of the toy model considered in \cite{Achucarro:2007qa}, where the heavy sector consisted of a single modulus. More precisely, there is always a basis such that the mass matrix and the Kähler metric can be diagonalized simultaneously. This allows expressing the stability requirement of having a positive definite mass matrix as a constraint on the curvature of the Kähler function at the uplifted vacuum (\ref{massmatrixev}).  We found that the stability diagram obtained for the toy model holds separately for each eigenvalue of the mass matrix of the uplifted scalar potential, fig. \ref{fig1}. In particular our results show that if the heavy fields are fixed at a minimum of the Kähler function the configuration remains stable for any final value of the cosmological constant. However, if the heavy fields are fixed at a saddle point of the Kähler function (the Kähler function cannot have maxima), the configuration always becomes unstable for large enough values of the cosmological constant. 

This analysis complements that of Covi et al. \cite{Covi:2008ea}, who formulated a necessary (and in most practical situations sufficient) condition for the existence of (meta)stable de Sitter vacua, following earlier work by Gomez Reino and Scrucca \cite{GomezReino:2006dk,GomezReino:2006wv,GomezReino:2007qi}. The constraint restricts the Kähler geometry of the non-linear sigma model associated to the chiral multiplets. Expressed in terms of the metric $G_{I\bar J}$ and the Riemann tensor $R_{I\bar J M\bar N}$ of the Kähler manifold  it reads:
\be 
  \sigma \equiv \biggl[ \frac{1}{3} \biggl(G_{I\bar J} G_{M \bar N} + G_{I\bar N} G_{M\bar J} \biggr) - R_{I\bar J M\bar N} \biggr] G^I G^{\bar J} G^M G^{\bar N} \label{sigma} > 0.
\ee
This condition, they point out, is e.g. not satisfied by moduli with no-scale Kähler functions of the form $K = -3 \log (\xi +\bar \xi) $, or more generally $K = -\sum_I n_I \log (\xi^I + \bar \xi^{\bar I}), \ \sum_I n_I = 3$. Clearly,  the constraint (\ref{sigma}) is only sensitive to the geometry of the Kähler manifold along the direction of the goldstino vector $G_I$,  and therefore it can say nothing about the perturbative stability of moduli with zero F-terms, $G_I=0$. In particular, it cannot be used to restrict the interactions of those fields that are supersymmetrically decoupled --in the sense of \cite{Achucarro:2008sy}-- from the sector that breaks supersymmetry.  Our work provides necessary and sufficient conditions for the perturbative stability of these $G^I = 0$ fields in a particular class of models where they are supersymmetrically decoupled.

Finally, we have also confirmed that the one-to-one correspondence found in \cite{Achucarro:2007qa} between local minima of the Kähler invariant function $G$ and (stable) AdS supersymmetric vacua that are local maxima of the scalar potential is completely general. These supersymmetric vacua satisfy the Breitenlohner-Freedman bound and are therefore stable. Our results imply that supersymmetric AdS maxima remain perturbatively stable when supersymmetry is broken by a supersymmetrically decoupled sector satisfying eq.  (\ref{ansatz}). Moreover, we have been able to prove that even in more general scenarios where the integrated heavy moduli do not satisfy (\ref{ansatz}), the supersymmetric AdS maxima are always stable for large values of the cosmological constant. To our knowledge, the uplifting of (AdS) supersymmetric local \emph{maxima} of the scalar potential has not been considered before and constitutes a new class of stable de Sitter vacua and inflationary troughs whose phenomenology has to be explored.

\acknowledgments{
  We thank Joe Conlon, Gonzalo Palma, Marta Gómez-Reino, Claudio Scrucca and Koenraad Schalm for useful discussions and comments.  Work supported by the Netherlands Organization of Scientific Research (N.W.O.)  under the VIDI and VICI programmes, by the Basque Government through grant BFI04.203 (K.S.) and by the Basque Government research project IT-357-07 (A.A.). We further acknowledge support by the Consolider-ingenio 2010 Programme CPAN (CSD2007-00042) and project FPA2005-04823.
}

\bibliography{literatuur}

\end{document}